# MOVING SIGNALS AND THEIR MEASURED FREQUENCIES

**Authors: Chandru Iyer[1] and *G. M. Prabhu[2]**

[1]Independent Researcher chandru_i@yahoo.com

[2]Department of Computer Science, Iowa State University, Ames, IA 50011, USA prabhu@iastate.edu

## ABSTRACT

Classical Doppler Effect of light propagation can be calculated by making any one of the two assumptions a) Light propagates at the speed c with respect to the source or b) Light propagates at the speed c with respect to the receptor. We show that the ratio of the two calculated classical Doppler Effects, with propagation speed c in the source inertial frame and receptor inertial frame respectively, remains constant equaling $[1-(v^2/c^2)]$ **:** 1 , in all orientations (longitudinal, transverse and any arbitrary angle). Furthermore, we show that the Doppler Effect as predicted by special relativity, is in each and every orientation, the geometric mean of the two expected classical Doppler Effects. This leads to two contradicting conclusions: (1) by the receptor that the clock associated with the emitter runs slow, and (2) by the emitter that the clock associated with the receptor runs slow. These differences can be resolved if we theorize that light travels at speed c with respect to the emitter as it leaves the emitter and travels at speed c with respect to the receptor as it approaches the receptor. The postulation that light travels at speed c with respect to material objects in the vicinity of those material objects will resolve all paradoxes related to reciprocal time dilation.

## INTRODUCTION

The Doppler Effect, originally proposed in 1842 by the Austrian physicist Christian Doppler, arises out of the difference in distance travelled by consecutive crests of a wave, while travelling from a source to a receptor, when the receptor and source are in relative motion [1]. It was further investigated by other scientists [2] in the $19^{th}$ century and is known by the name of Doppler as he was the first to propose the existence of the phenomenon in a theoretical framework. The Doppler Effect is widely used in medical scanning applications [3]. It is physically perceptible to the normal human faculties in the case of an approaching or leaving source of sound. In classical physics the Doppler Effect can be calculated by knowing:

- the orientation of the line joining the source and receptor with respect to the line of relative motion of the source and receptor,
- the emitting frequency, and
- the speed of propagation with respect to the source.

In the receptor's assessment, the movement of the emitter during the time of propagation of the signal from emitter to receptor does not figure in the relevant calculations. In the emitter's assessment, the movement of the receptor during the time of propagation of the signal from emitter to receptor is relevant and significant. This perception gives rise to the observed distances travelled by a signal between emitter and receptor to be different as observed by the emitter and receptor respectively. The difference in distances is compensated, in classical physics, by the difference in speed of propagation of the signal as observed by the two inertial frames associated with the source and receptor, and one calculates identical results from either inertial frame. The expected Doppler Effect under classical physics calculated by the receptor and emitter are different, when one assumes that the speed of light is the same with respect to both inertial frames. Any set of observations with associated space-time coordinates is essentially classical as far as observers in one inertial frame are concerned. Thus these observers will interpret the observations under classical physics. It is in this context that such observers observe that moving clocks run slow, assuming that the propagation of light is at constant speed whether the source of light is stationary or moving.

However, the orientation is dynamic and as the relative motion progresses, the orientation continually changes. Hence the Doppler Effect also changes as the relative motion progresses. This is the reason that the whistle of an approaching train is shriller to a person on the ground. When the train is far away, the orientation angle between the line of motion and the line joining the receptor and the whistle is almost zero. It is never exactly zero unless the receptor is a point on the line of motion, which is not the case as the person on the ground will not usually position himself along the line of motion. As the train approaches closer the orientation shifts from zero to 180 degrees in a continuous fashion (see Fig. 1).

The Doppler Effect depends on the orientation [4]. This is why an approaching siren is shriller and slides to actual frequency as the siren passes you (orientation $90^0$) and becomes low pitched as the siren leaves the observer. In principle, between two subsequent (emissions of) crests the orientation changes slightly. But this change can be neglected when the frequency is very high compared to the reciprocal of the time taken by the wave to travel between source and receptor. When the orientation is $90^0$ and 'd' in Fig. 1 is very small, this change in orientation, that is, the change in $\theta$ between two consecutive crests, cannot be neglected.

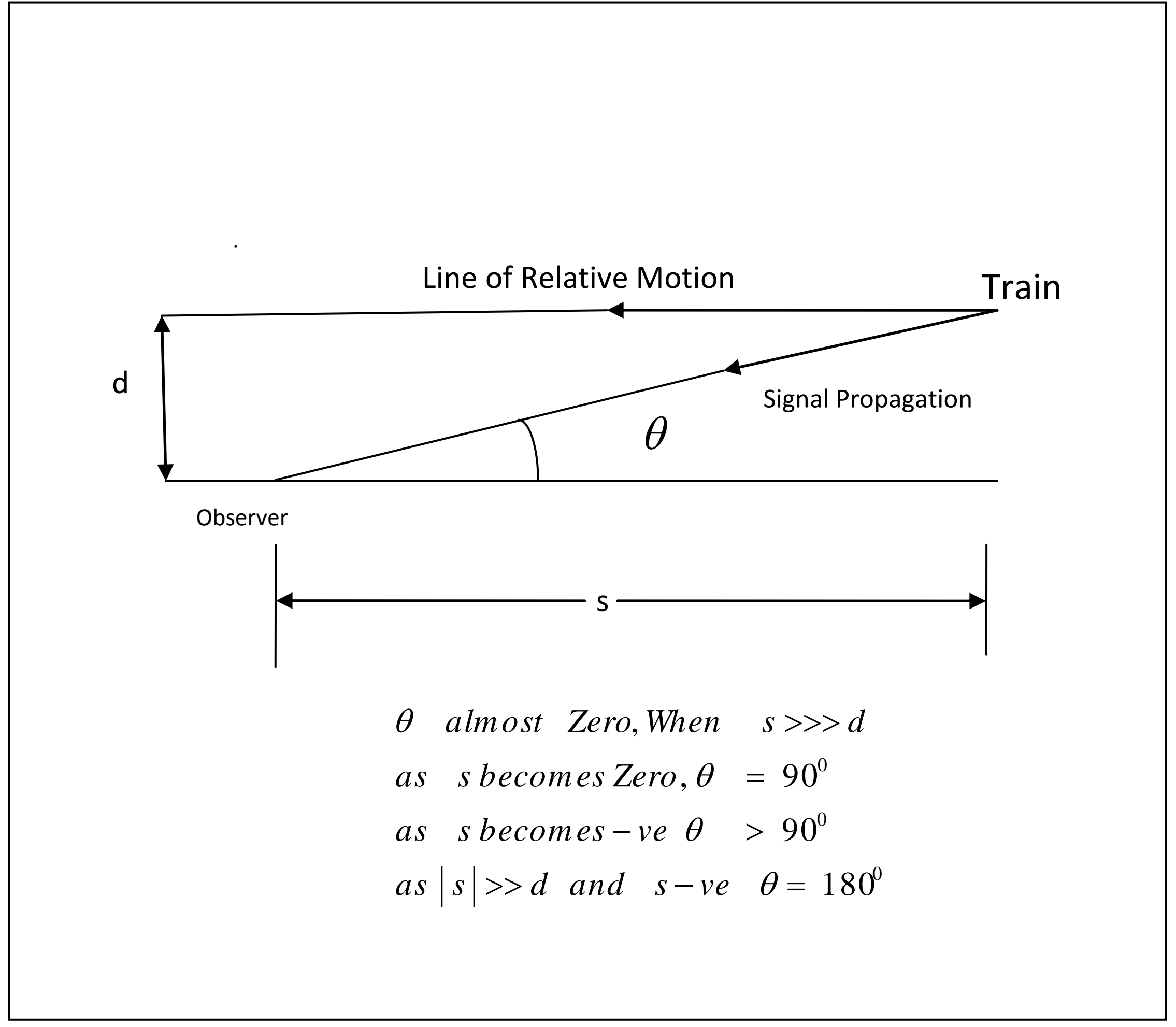

Figure 1. Doppler Effect and dynamic orientation

In classical physics the orientation θ, at any given instant, is the same as observed by the receptor and source. This is because the definition of an instant of time is same in both frames and both agree on a snapshot of events at a given instant as the same. In relativistic physics, the two inertial frames do not agree on an instant and thus they disagree on the orientation at any instant of emission of a crest. The disagreement in the definition of the instant arises from the long debated convention in defining simultaneity that is elaborated in its historical, scientific and philosophical aspects by Janis [5]. In relativistic physics the location of the receptor with respect to the source, when a crest is emitted by the source, is not the same as observed by the inertial frame co-moving with the receptor and as observed by the inertial frame co-moving with the source. Since the Doppler Effect depends on orientation, this perception of difference in orientation leads to different results as calculated by observers in both the inertial frames. Moreover both sets of observers assume that light travels at a speed $c$ with respect to the inertial frame that they are at rest. We show that the observed Doppler Effect is the geometric mean of the two calculated (classical) ones by the two sets of observers.

Table 1 enumerates the physical aspects that contribute to the Doppler Effect in classical and relativistic physics.

|  | **Relativistic Physics** | **Classical Physics** |
|---|---|---|
| 1. | There is a contribution to the perception of difference in distance travelled by a signal (crest) as observed by Receptor and Source on account of the perceptions that Receptor moves (according to source) and the Receptor is stationary (according to Receptor) during the propagation of the wave/ signal | In agreement that the Receptor moves (according to Source) and the Receptor is stationary (according to Receptor) during the propagation of the wave/ signal. In agreement that the distance travelled by the crest is observed to be different by the Receptor and Source because of these different perceptions. |
| 2. | Speed of propagation (of light) is identical as observed by Source and Receptor. | Speed of propagation as observed by the Source and Receptor are different and differ exactly by the relative velocity between the Source and Receptor. This compensates for the effect in number1 above, resulting in identical results (Doppler Effect) as calculated by the two inertial frames associated with the Source and Receptor. |
| 3. | Distances travelled by the signal are different as observed by the two inertial frames because of the differences in perception of the instant itself. | There is agreement in what is perceived as an instant and therefore no difference in travel distance is attributable on this account. |
| 4. | Instant of Receipt is different in both the frames as the source and receptor are spatially separated and they cannot agree on the instant snapshot of the physical world. (Remarks: (a) Instant of emission is synchronized between the two frames as $t = 0$; $t' = 0$. (b) If the instant of receipt is synchronized between the two frames, then the instant of emission is perceived to be different!) | The Source and Receptor agree on the instant of time and no differences arise on this account ($t' = t$). |
| 5. | Location of receipt is perceived / observed to be different (if the location of emission is taken as $x = 0$; $x' = 0$). Location of Emission is perceived / observed to be different (if the location of receipt is taken as $x = 0$; $x' = 0$). | Location of Receipt is observed to be different only to the extent of the relative motion ($x' = x-vt$). This is accounted for in Number 1 as above and compensated by Number 2 as above. Thus both the inertial frames agree on the calculated Doppler Effect. However, if both the inertial frames assume the speed of propagation as same and identical (as c), then the difference created by Number 1 is not compensated and both the inertial frames calculate the Doppler Effect to be different. |
| 6. | In relativistic physics, the distance between the source and the receptor at the instant of emission of the signal is observed to be different by the two inertial frames, arising out of the disagreement on the concept of instant itself due to asynchronization of spatially separated objects/clocks (source and receptor). So all calculations based on constructive methods such as time at instant of signal emission + time duration of travel of signal from source to receptor = time at instant of receipt, cannot be used in a simple manner especially when two consecutive signals, separated by a time interval (reciprocal of frequency), are originating from the source and reaching the receptor. So the relativistic Doppler Effect is calculated by using the Lorentz transformations and not by the equation<br>Time at instant of signal emission + Time duration of travel of signal from source to receptor = Time at instant of receipt.<br>This is because there is disagreement on the instant of emission and instant of receipt between the two inertial frames. | In classical physics, the distance between the source and the receptor at the instant of emission of the signal is observed to be the same by the two inertial frames. The instant of receipt is also observed to be the same. Only the distance travelled and the propagation speed are observed to be different; but they compensate each other resulting in the time taken to travel from the source to receptor to be identical. Once we assume the propagation speed to be same, the calculated Doppler Effects by both the frames become different. The relativistic Doppler Effect is the geometric mean of these two classical Doppler Effects in all geometries and orientations. In all cases, the classical Doppler Effect can be calculated from first principles, that is,<br>Time at instant of signal emission + Time duration of travel of signal from source to receptor = Time at instant of receipt. |

Table 1. Factors that affect Doppler Effect in classical and relativistic physics

With the above considerations in mind, it is possible to constructively calculate the Doppler Effect only in the case of classical physics. We may use either inertial frame to calculate the Doppler Effect, but we need to know the speed of propagation with respect to that inertial frame.

When we calculate the classical Doppler Effect by two choices, namely, (a) based on observations by observers co-moving with the source, and (b) based on observations by observers co-moving with the receptor, the two results are identical only when the speed of propagation assumed in (a) and the speed of propagation assumed in (b) differ exactly by the relative velocity between the inertial frames. Therefore, in one of the inertial frames, the speed of propagation is direction dependent (anisotropic) whereas in the other frame the speed of propagation is isotropic.

As a consequence when we assume both the frames to be isotropic, the calculated Doppler Effects in cases (a) and (b) are not identical. Therefore, one may state that the classical Doppler Effect gives two separate results: one by assuming that the inertial frame associated with the source is isotropic, and the other by assuming that the inertial frame associated with the receptor is isotropic. In this context when we say an inertial frame is isotropic, we mean that the propagation of light from one location to another location within that inertial frame is at speed c in all directions irrespective of whether the light source is stationary or moving with respect to that inertial frame.

The relativistic Doppler Effect is calculated by using the Lorentz Transformations in standard texts [6, 7, 8]; a constructive derivation is precluded by the relativistic concept of 'relativity of simultaneity.' Since each inertial frame adopts its own convention of simultaneity [5], the instant at which receipt of a signal takes place is observed to be different and the time interval between emission and receipt is observed to be of different durations by the inertial frames associated with the rest frames of the source and the receptor.

In this paper we formulate the two classical Doppler effects (1) by assuming that the inertial frame associated with the source is isotropic, and (2) by assuming that the inertial frame associated with the receptor is isotropic. We develop formulations for (1) and (2) in the orientations longitudinal, transverse, and arbitrary orientation and show that the ratio of the observed (by the receptor) frequencies calculated in (1) and (2) remain constant in all the orientations. Further, we show that the relativistic Doppler Effect is the geometric mean of the two classical Doppler Effects in (1) and (2) in all orientations. We also present some conclusions from the fact that the observed Doppler Effect is the mean of the two classical Doppler Effects.

## Classical Longitudinal Doppler Effect: Light speed is c with respect to Source

Consider (as illustrated in Fig. 2) a source of light moving at speed $v$ with respect to the laboratory frame of reference. Assume also that light is emanating from the source travelling in all directions at speed $c$ with respect to the source. The speed of the light rays with respect to the Laboratory reference frame is to be evaluated by the Galilean / Newtonian formulas for additive velocities, as done in classical physics.

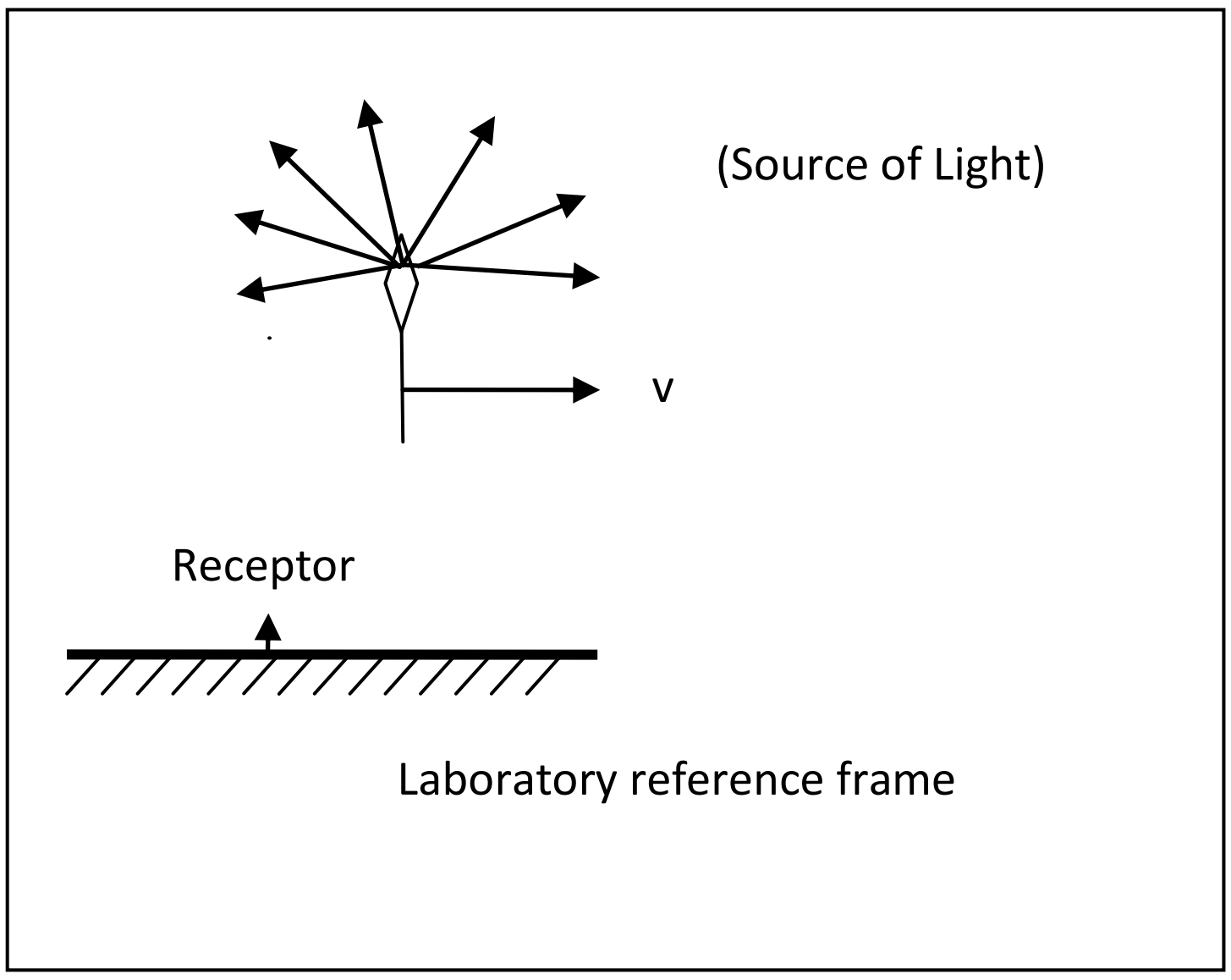

Figure 2. Source moving with respect to laboratory reference frame

We use primed notation for the inertial frame co-moving with the source (K′) and unprimed notation for the 'stationary' laboratory reference frame (K) in which the receptor is situated.

This is consistent with the way the subject matter is dealt with in standard texts [6] wherein it is conceived that a moving light source is observed by a 'stationary' reference frame in which a laboratory with necessary instruments is located.

The frequency $f'$ of the source of light is the number of peaks (of wave fronts) emanating per second. The two consecutive peaks may be considered as two signals separated by a time interval $(1/f')$. For a light wave, this frequency is of the order of $10^{15}$ and thus it is a very large number and in our evaluations we will be justifiably neglecting quantities of the order of magnitude $1/(f')^2$. We use $f'$ as the frequency of the source of light consistent with the assumption that a 'moving' source (primed frame) is being observed by a 'stationary' laboratory.

For the longitudinal Doppler Effect, we consider only events on the x- axis (Fig. 3). In this case, the line joining the source and receptor is collinear with the line of relative motion.

R S v
Receptor Source

Figure 3. Source moving away from receptor on x-axis

Assume that the source of light is moving with respect to the instrument receiving the signals at a speed $v$ and away from the instrument. Let us also say that at a given instant t = 0, the distance between the source and the instrument is $s$. At this instant the first signal emanates from the source travelling at speed $c$ with respect to the source and at speed $(c-v)$ with respect to the instrument, in accordance with classical physics.

The first signal reaches the instrument at $t = \dfrac{s}{c-v}$ .

At the instant $\dfrac{1}{f'}$ , the second signal is emanated from the source and at this instant the separation between the source and instrument is $s + \dfrac{v}{f'}$ and the signal starts at $t = \dfrac{1}{f'}$ and reaches the instrument after a time interval of $\dfrac{s + \frac{v}{f'}}{c-v}$ and the instant it reaches the instrument is $\dfrac{1}{f'} + \dfrac{s + \frac{v}{f'}}{c-v}$ .

Therefore the frequency $f$ measured by the Laboratory reference frame is derived as follows:

$$\frac{1}{f} = \frac{1}{f'} + \frac{s + \frac{v}{f'}}{c-v} - \frac{s}{c-v} \qquad \text{--------------- (1)}$$

$$\frac{1}{f} = \frac{1}{f'} + \frac{v}{c-v} \cdot \frac{1}{f'} \qquad \text{-------------- (2)}$$

$$f = f'\left(1 - \frac{v}{c}\right) \qquad \text{-------------- (3)}$$

Equation (3) gives the classical longitudinal Doppler Effect when the speed of light is *c* with respect to the source.

## Classical Longitudinal Doppler Effect: Light speed is c with respect to Receptor

For this case we refer to Fig. 3, but we will take the speed of the light ray to be *c* with respect to the receptor.

Two signals or two consecutive peaks are separated by a time interval ($1/f'$) in the reference frame co-moving with the source. Assume that the source of light is moving with respect to the instrument receiving the signals at a speed of *v* and away from the instrument. Let us also say at a given instant t = 0, the distance between the source and the instrument is *s*. At this instant the first signal emanates from the source travelling at speed *c* with respect to the laboratory reference frame. The instant at which this signal reaches the instrument is $\frac{s}{c}$. At the instant $\frac{1}{f'}$ the second signal is emanated from the source and at this instant the separation between the source and the instrument is $s+\frac{v}{f'}$. The instant at which this signal reaches the instrument is $\frac{1}{f'}+\frac{s+\frac{v}{f'}}{c}$.

Therefore, the frequency $f$ measured by the laboratory reference frame is derived as follows:

$$\frac{1}{f}=\frac{1}{f'}+\frac{s+\frac{v}{f'}}{c}-\frac{s}{c} \qquad \text{----------------- (4)}$$

$$\frac{1}{f}=\frac{1}{f'}+\frac{v}{c}.\frac{1}{f'} \qquad \text{----------------- (5)}$$

$$f=\frac{f'}{1+\frac{v}{c}} \qquad \text{---------------- (6)}$$

Equation (6) gives the classical longitudinal Doppler Effect when the speed of light is *c* with respect to the receptor. In both equations (3) and (6) the source and the receptor are moving away from each other.

The relativistic Doppler Effect is:

$$f=\frac{f'\sqrt{1-\frac{v}{c}}}{\sqrt{1+\frac{v}{c}}} = f'\sqrt{\frac{c-v}{c+v}} \qquad \text{-------------- (7)}$$

(see also Equation 2 – 29, page 90 of [6]).

We observe that the relativistic Doppler Effect is the geometric mean of the two longitudinal Doppler Effects given by equations (3) and (6) respectively.

We now proceed to show that the relativistic transverse Doppler Effect is also the geometric mean of

1) the classical transverse Doppler Effect when the speed of light is *c* with respect to the receptor and

2) the classical transverse Doppler Effect when the speed of light is *c* with respect to the source.

## Classical Transverse Doppler Effect: Light speed is c with respect to Receptor

The classical transverse Doppler Effect is when the Laboratory instrument receives the signal at $90^0$ with respect to the line of motion of the source. For the case when the speed of light is *c* with respect to the receptor we may refer to Fig. 4.

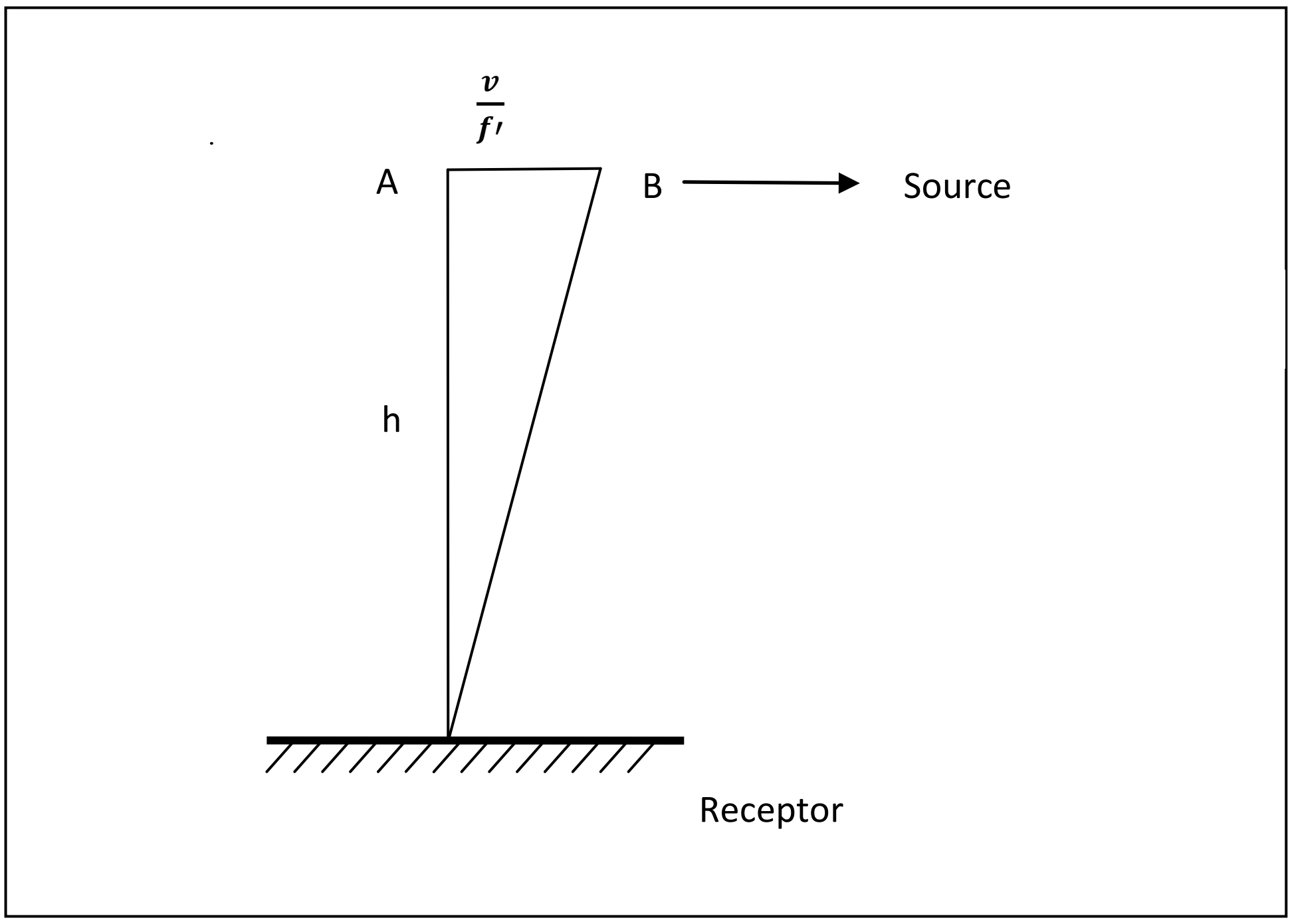

Figure 4. Signal received perpendicular to line of motion

Point A signifies the emanation of the first signal at t = 0. Point B signifies the emanation of the second signal at $t = \dfrac{1}{f'}$. The distance AB = $\dfrac{v}{f'}$ .

The first signal reaches the receptor at $t_1 = \frac{h}{c}$; the second signal reaches the receptor at

$t_2 = \frac{1}{f'} + \frac{\sqrt{h^2 + \frac{v^2}{f'^2}}}{c}$ and therefore the frequency observed by the receptor is given by

$$\frac{1}{f} = t_2 - t_1 = \frac{1}{f'} + \frac{\sqrt{h^2 + \frac{v^2}{f'^2}}}{c} - \frac{h}{c} \qquad \text{----------------} \ (8)$$

Considering that h and $f'$ are large, the above equation reduces to

$$\frac{1}{f} = \frac{1}{f'} \qquad \text{-----------------} \ (9)$$

$$\text{Or} \quad f = f' \qquad \text{------------------} \ (10)$$

This is referred to as the inability of classical physics to predict a transverse Doppler effect [6]. However, we will show in the next section that classical physics does predict a transverse Doppler effect when the speed of light is *c* with respect to the source.

## Classical Transverse Doppler Effect: Light speed is c with respect to Source

In the case when the speed of light is *c* with respect to the source, the transverse Doppler Effect can be visualized as shown in Fig. 5.

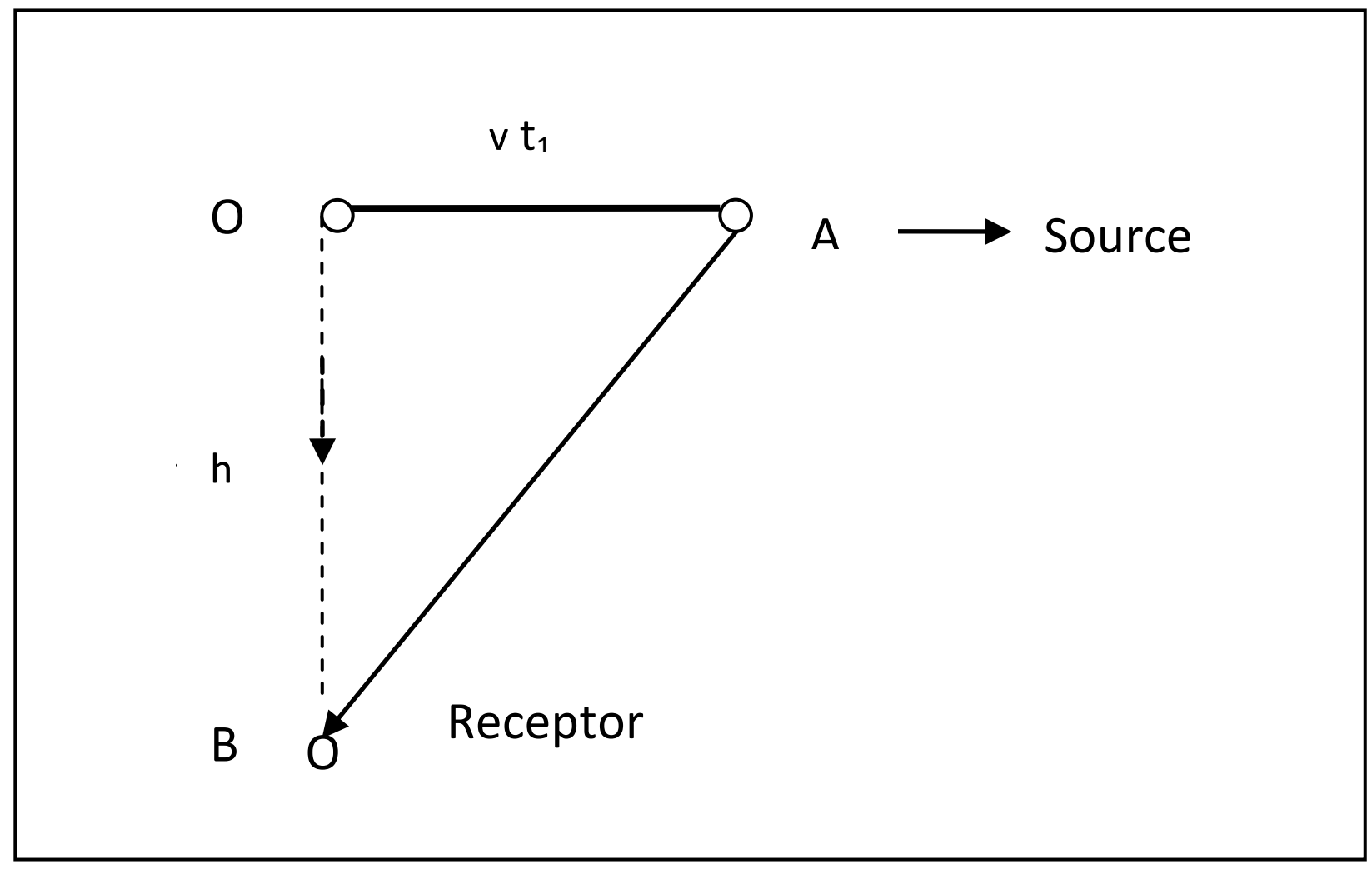

Figure 5. Signal received perpendicular to line of motion

At t = 0, a peak of the wave front of light (the first signal) emanates from the source and reaches the receptor B at t = $t_1$. At this instant t = $t_1$, the source has moved and is now (at t = $t_1$) located at point

A, and the distance travelled by the light ray is AB from the source. Therefore (AB/c) = $t_1$, when the light is travelling at **c** with respect to the source of light. (Please note that all observers stationary with respect to the inertial frame co-moving with the Receptor observe that the light travels along the line OB, whereas all observers stationary with respect to the inertial frame co-moving with the source of light observe that the light ray travels along the line AB).

In order to determine the classical transverse Doppler Effect when the speed of light is *c* with respect to source, it is recommended to have the reference frame co-moving with the source as the preferred reference frame. This facilitates the easy determination of distances travelled by the light signals from the source as observed by the frame co-moving with the source. This is depicted in Fig. 6.

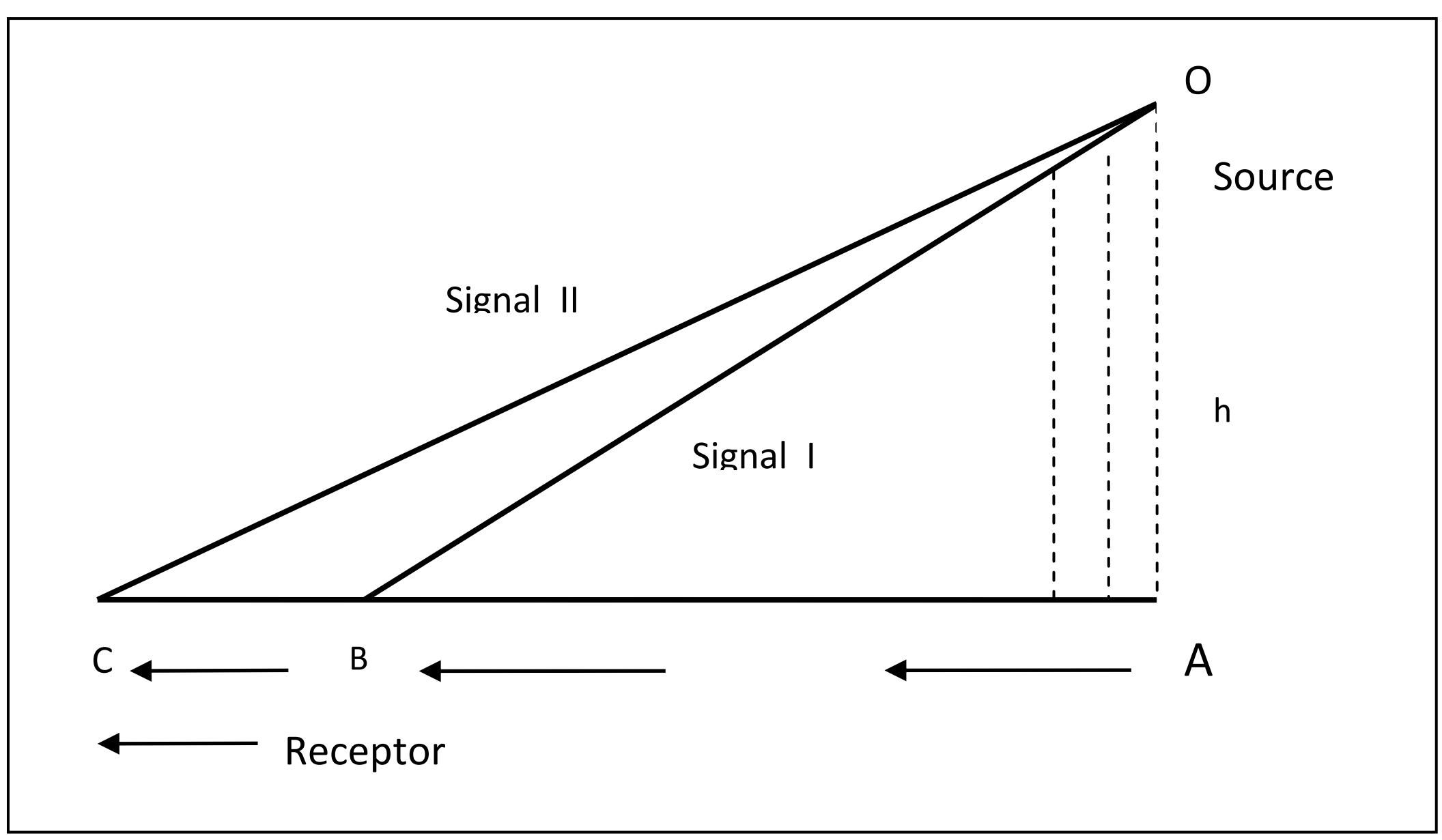

Figure 6. Speed of light is *c* with respect to the source

To be consistent with our formulations in previous sections where we had the receptor stationary and the source moving at *v* along positive x axis, we now have the source stationary and the receptor moving at –*v*.

Point O is the source. At time $t_1 = 0$ signal I (a peak in the wave front) emanates from the source. At this instant the receptor is located at A. As the light signal moves from O to B, the receptor moves from A to B. The first signal is received by the receptor at location B. Evidently, if $t_1$ is the time at which signal I is received by the receptor, then

$$AB = v\,t_1 \qquad \text{-------------- (11)}$$

$$OB = ct_1 \qquad \text{-------------- (12)}$$

$$\angle BAO = 90^0 \qquad \text{------------------------(13)}$$

The dotted lines indicate that, as observed by the receptor, the light signal OB travels perpendicular to the line of relative motion.

The second light signal emanates from O at $t_2 = \frac{1}{f'}$; $f'$ being the frequency of the source light beam. This signal reaches point C at time $t_2$.

Evidently AC = $v$ $t_2$ and the frequency observed by the receptor is given by $\frac{1}{f} = t_2 - t_1$.

$$t_1 = \frac{OB}{c} = \frac{\sqrt{h^2 + v^2 t_1^2}}{c} \quad \text{------------ (14)}$$

$$t_2 = \frac{OC}{c} + \frac{1}{f'} = \frac{\sqrt{h^2 + v^2 t_2^2}}{c} + \frac{1}{f'} \quad \text{------------ (15)}$$

$$\left( \frac{OC}{c} \text{ being the time elapsed; } \frac{1}{f'} \text{ being the start time} \right)$$

$$t_2 - t_1 = \frac{1}{f} = \frac{1}{f'} + \frac{\sqrt{(h^2 + v^2 t_2^2)}}{c} - \frac{\sqrt{(h^2 + v^2 t_1^2)}}{c} \quad \text{----------- (16)}$$

By applying Pythagoras theorem for triangle OAB , we get

$$h = \sqrt{\left(c^2\ t_1^2 - v^2 t_1^2\right)} = c t_1 \sqrt{1 - \frac{v^2}{c^2}} \quad \text{------------ (17)}$$

By applying Pythagoras theorem for triangle OAC, and noting from equation (15) that OC = $c$[$t_2$ – (1/$f'$ )] we get

$$h = \sqrt{\left[ c^2 \left( t_2 - \frac{1}{f'} \right)^2 - v^2 t_2^2 \right]} \quad \text{---------- (18)}$$

Therefore, from equations (17) and (18) we obtain

$$c^2 \left( t_2 - \frac{1}{f'} \right)^2 = c^2\ t_1^2 - v^2\ t_1^2 + v^2\ t_2^2 \quad \text{-------------- (19)}$$

$$\left( t_2 - \frac{1}{f'} \right)^2 = t_1^2 - \frac{v^2}{c^2}\ t_1^2 + \frac{v^2}{c^2}\ t_2^2 \quad \text{-------------- (20)}$$

$$t_2^2 - \frac{2 t_2}{f'} + \frac{1}{f'^2} = t_1^2 - \frac{v^2}{c^2}\ t_1^2 + \frac{v^2}{c^2}\ t_2^2 \quad \text{-------------- (21)}$$

The frequency of light being normally a very large number, $\frac{1}{f'^2}$ is negligible, being a square of the time difference between two signals (peaks of the wave front). Therefore,

$$t_2^2 - \frac{2t_2}{f'} = t_1^2 - \frac{v^2}{c^2} t_1^2 + \frac{v^2}{c^2} t_2^2 \quad \text{-------------------- (22)}$$

$$t_2^2 \left(1 - \frac{v^2}{c^2}\right) = t_1^2 \left(1 - \frac{v^2}{c^2}\right) + \frac{2t_2}{f'} \quad \text{-------------------- (23)}$$

$$t_2^2 = t_1^2 + \frac{2t_2}{f'} \frac{1}{1 - \frac{v^2}{c^2}} \quad \text{-------------------- (24)}$$

$$t_2^2 - [\frac{2t_2}{f'} \frac{1}{1 - \frac{v^2}{c^2}}] = t_1^2 \quad \text{-------------------- (25)}$$

By completing the square on the LHS by adding and subtracting the term $1/[f'^2 (1-v^2/c^2)^2]$, we obtain

$$\left[ t_2 - \frac{1}{f'\left(1 - \frac{v^2}{c^2}\right)} \right]^2 - \frac{1}{f'^2 \left(1 - \frac{v^2}{c^2}\right)^2} = t_1^2 \quad \text{-------------------- (26)}$$

The second term on the LHS can be ignored as $\frac{1}{f'^2}$ is a second order term of small quantity (since $f'$ is large).

$$\therefore t_2 - \frac{1}{f'\left(1 - \frac{v^2}{c^2}\right)} = t_1 \quad \text{-------------------- (27)}$$

$$t_2 - t_1 = \frac{1}{f'\left(1 - \frac{v^2}{c^2}\right)} \quad \text{-------------------- (28)}$$

or $$\frac{1}{f} = \frac{1}{f'\left(1 - \frac{v^2}{c^2}\right)} = (\gamma^2 / f') \quad \text{-------------------- (29)}$$

where we follow the usual notation $\gamma^2 = 1/ [1-(v^2/c^2)]$ ---------------------(30)

Therefore, $$f = \frac{f'}{\gamma^2} \quad \text{-------------------- (31)}$$

Equation (31) $f = \frac{f'}{\gamma^2}$ gives the classical transverse Doppler Effect when the speed of light is $c$ with respect to the source.

Equation (10) $f = f'$ gives the classical transverse Doppler Effect when the speed of light is $c$ with respect to the receptor.

The relativistic transverse Doppler Effect is $f = \frac{f'}{\gamma} = f'\sqrt{1-\frac{v^2}{c^2}}$ -------------------- (32)

(see also Eqn 2-30, page 90 of.[6]).

Thus the relativistic transverse Doppler Effect as given by equation (32) is the geometric mean of the two classical Doppler effects given by equations (10) and (31) .

## Doppler Effects at arbitrary angle of emission and receipt

From the preceding four sections, we can see that for both the longitudinal and transverse cases the ratio of the three quantities – "Doppler effect predicted by classical theory with speed of light as **c** with respect to source": "Doppler effect predicted by relativity theory" : "Doppler effect predicted by classical theory with speed of light as **c** with respect to receptor" remains (1/ γ) : 1 : γ. In this section we show that the ratio of these three quantities remains (1/ γ): 1: γ for all angles of emission and receipt.

<u>Speed of light is c with respect to source</u>

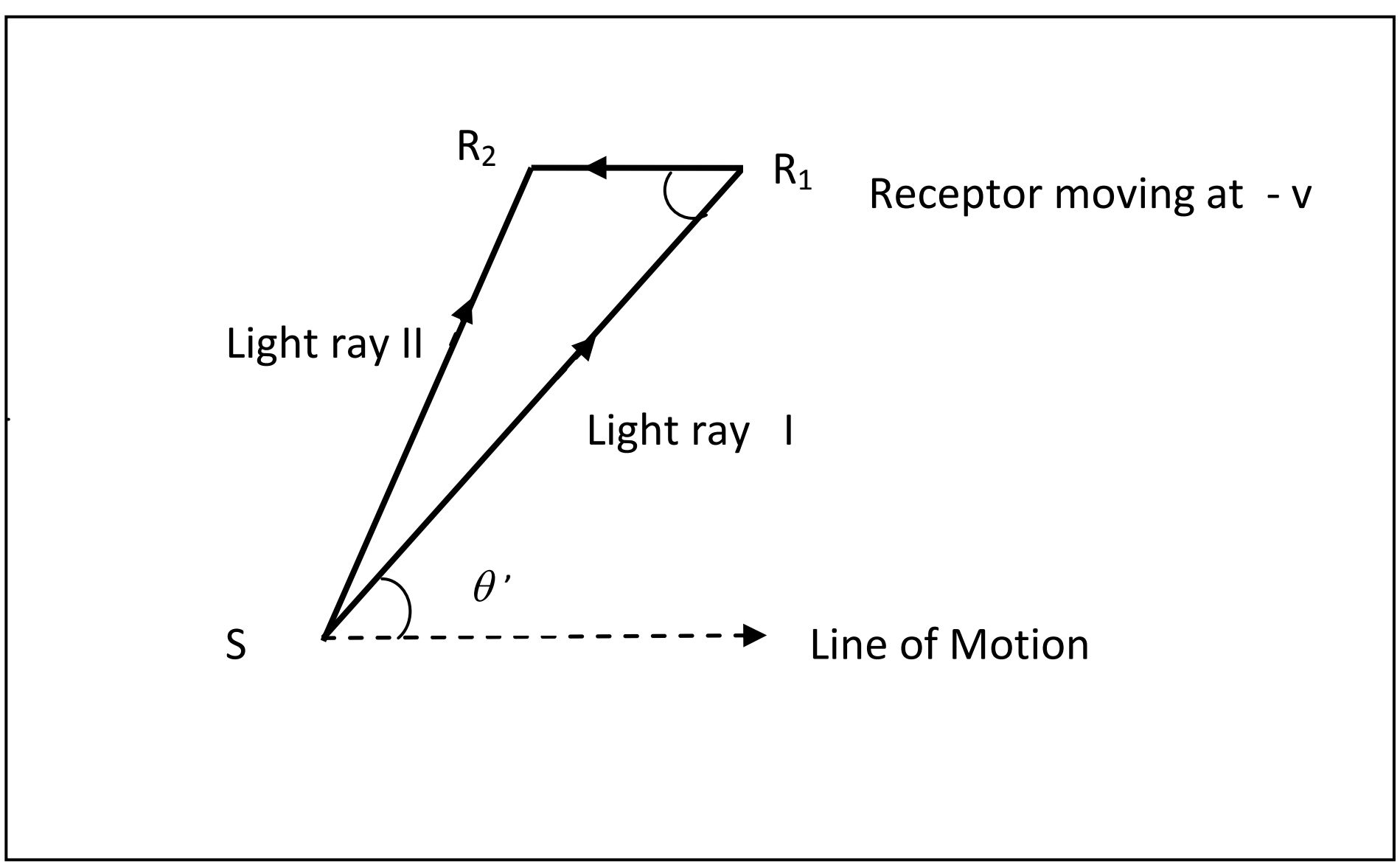

Figure 7. Observations by inertial frame co-moving with source (speed of light $c$ with respect to source)

S is the source, $R_1$ is the position of receptor at receipt of first signal, and $R_2$ is the position of receptor at receipt of second signal. $\theta'$ is the angle at which light ray is emitted from source with reference to the line of motion (as observed by inertial frame co-moving with source).

The primed notation $\theta'$ is used consistent with formulation of the problem that the source is moving.

Further we have designed the diagram (Fig. 7) in such a way that $\theta'$ is acute and positive so that sign conventions of trigonometric functions are automatically taken care of.

By the formulations in geometry we have

$$SR_2^2 = SR_1^2 + (R_1 R_2)^2 - 2(SR_1)(R_1 R_2) Cos\ \theta'$$

With frequency of light being very large, $R_1 R_2$ is very small. The square of $R_1 R_2$ may be neglected and we have

$$SR_2^2 = SR_1^2 - 2(SR_1)\ (R_1 R_2) Cos\ \theta'$$

$$= (SR_1 - R_1 R_2\ Cos\ \theta')^2 - (R_1 R_2 Cos\ \theta')^2$$

The last term can be neglected as $R_1 R_2$ is very small.

Therefore we have $SR_1 = SR_2 + R_1 R_2\ Cos\theta'$

Therefore the second signal takes a little less time in reaching the receptor given by $\frac{R_1 R_2\ Cos\theta'}{c}$. Further the second signal leaves a little later given by the quantity $\frac{1}{f'}$.

Therefore the frequency of receipt $f$ is given by the equation

$$\frac{1}{f} = \frac{1}{f'} - \frac{R_1 R_2\ Cos\theta'}{c}$$

Now $R_1 R_2 = \frac{v}{f}$ ; the distance travelled between the two receptions.

Therefore $\frac{1}{f} = \frac{1}{f'} - \frac{v}{f\ c} \quad Cos\theta'$

$$\text{or} \quad f = f'\left(1 + \frac{v}{c}\cos\theta'\right) \qquad \text{-------------} \quad (33)$$

Equation (33) gives the classical Doppler Effect at arbitrary angle of emission θ′ when the speed of light is *c* with respect to the source. *f* is the expected receptor frequency and $f'$ is the source frequency.

Comparing with the relativistic prediction given in [6] equation 2-25a pp89,

$$f = f'\left(1 + \frac{v}{c} Cos\theta'\right)\gamma \qquad \text{Reference [6] equation 2-25a.}$$

Comparing the above with equation (33) we have the ratio of the classical prediction with light travelling at *c* with respect to the source to the relativistic prediction to be (1/ γ ) : 1.

Speed of light is c with respect to receptor

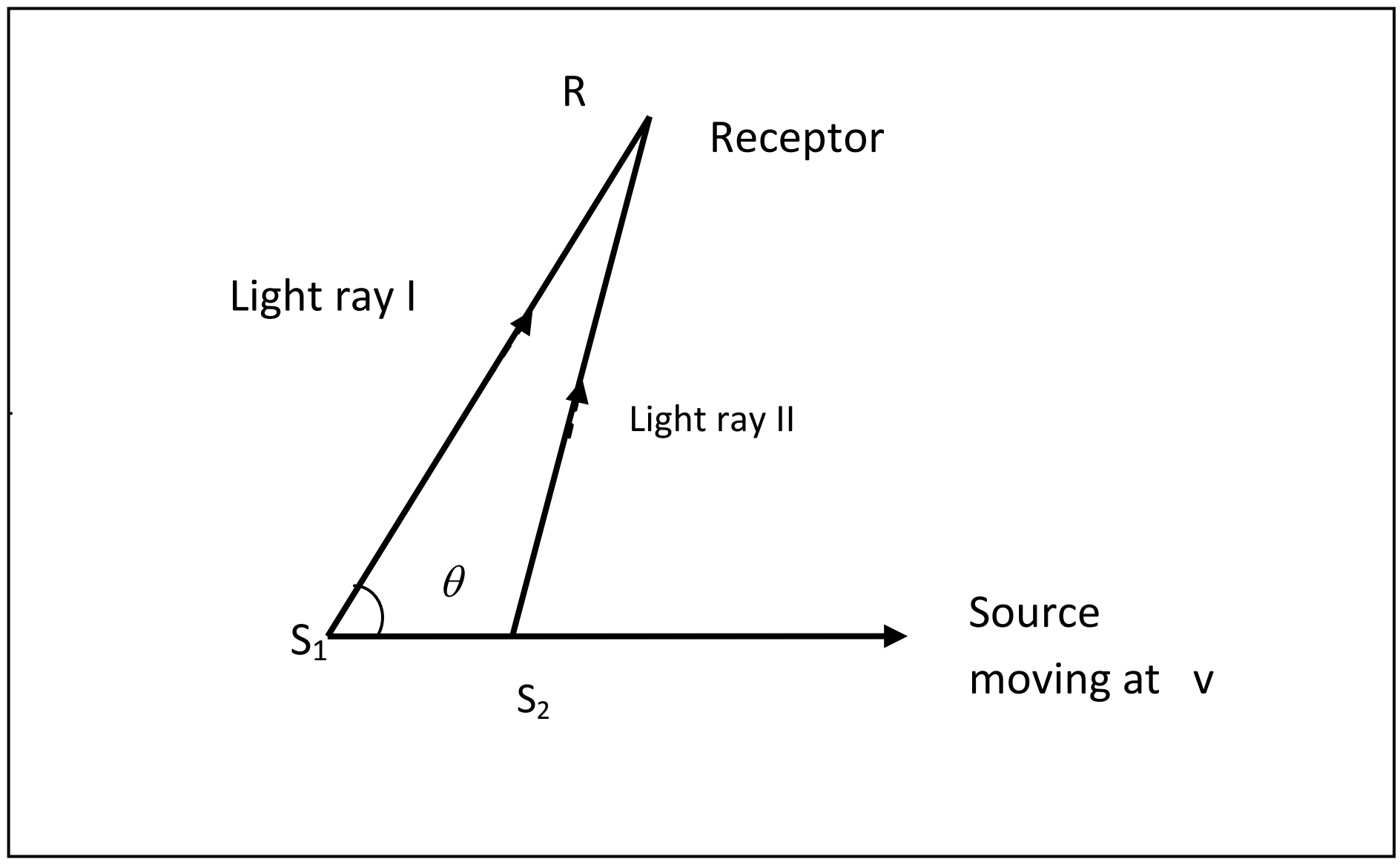

Figure 8. Observations by inertial frame co-moving with receptor (speed of light *c* with respect to receptor)

R is the receptor. $S_1$ is the position of the source at the instant of emission of first signal. $S_2$ is the position of the source at the instant of emission of second signal. $\theta$ Is the angle at which light ray is emitted from source with reference to the line of motion (as observed by the inertial frame co-moving with receptor). The notation $\theta$ (not primed), is used for the angle observed by the inertial frame co-moving with the receptor, consistent with the formulation of the problem that the receptor is at rest and source is moving.

Moreover we have designed the diagram (Fig. 8) in such a way that $\theta$ is acute and positive so that sign conventions of trigonometric functions are automatically taken care of.

By the formulations in geometry we have $(S_2 R)^2 = (S_1 R)^2 + (S_1 S_2)^2 - 2(S_1 R)(S_1 S_2) Cos\ \theta$

Neglecting $(S_1 S_2)^2$ as $S_1 S_2$ is very small we have

$$(S_2 R)^2 = (S_1 R)^2 - 2(S_1 R)\ (S_1 S_2)\ Cos\ \theta$$

or $(S_2 R)^2 = \left[S_1 R - (S_1 S_2) Cos\theta\right]^2 - (S_1 S_2 Cos\theta)^2$

Neglecting the last term as $S_1$ $S_2$ is very small we have $S_1 R = S_2 R + (S_1 S_2)\ Cos\theta$ .

Therefore signal II reaches receptor faster but it is emanated after a time lag of $\frac{1}{f'}$ the reciprocal of the source frequency.

Therefore, the observed frequency is given by $\frac{1}{f} = \frac{1}{f'} - \frac{S_1 S_2 Cos\theta}{c}$

The term $\frac{S_1 S_2 \, Cos\theta}{c}$ is the time difference due to the shortening of the travel path. Now

$S_1 S_2 = \frac{v}{f'}$ that is the distance travelled between two consecutive signals emanating from the source.

$$\text{Therefore} \quad \frac{1}{f} = \frac{1}{f'} - \frac{v \; Cos\theta}{f'c}$$

$$or \quad f = \frac{f'}{1 - \frac{v}{c} \; Cos\theta} \qquad \text{--------------} \quad (34)$$

Equation (34) gives the classical Doppler Effect at arbitrary angle of receipt θ when the speed of light is *c* with respect to the receptor. *f* is the observed receptor frequency and *f '* is the (calculated) source frequency.

Comparing with the relativistic prediction given in Equation 2- 25b pp 89 of reference [6]

$$f' = \frac{f\left(1 - \frac{v}{c} Cos\theta\right)}{\sqrt{1 - (v^2 / c^2)}} \qquad \text{Reference [6] equation 2-25b}$$

which can be re-written as $f = \frac{f'\sqrt{1 - \frac{v^2}{c^2}}}{1 - \frac{v}{c} Cos\theta}$ Reference [6] equation 2-25b

(rearranged)

$$= \frac{f'}{1 - \frac{v}{c} Cos\theta} \quad \frac{1}{\gamma} \qquad \text{Reference [6] equation 2-25b}$$

(rearranged)

Comparing the relativistic prediction with the one predicted by classical physics (equation 34), the ratio of the relativistic prediction of the frequency of receipt to the classical prediction with light travelling at *c* with respect to the receptor is $1 : \gamma$

Thus we have for all angles of emission and receipt, the ratio of the three quantities, namely,

"frequency of receipt predicted by classical physics with speed of light *c* with respect to source": "relativistic prediction of frequency of receipt": "frequency of receipt predicted by classical physics with speed of light *c* with respect to receptor" = $\left(\frac{1}{\gamma}\right) : 1 : \gamma$ . ----- (35)

## CONCLUSION

There are two possible interpretations for the ratio described by Equation (35) in the previous section.

**Relativistic Interpretation:** The relativistic interpretation of the ratio is as follows. Observers co-moving with the source expect the receptor to observe a particular frequency, whereas the receptor observes a higher frequency increased by the factor $\gamma$. This is interpreted by the observers co-moving with the source due to the slow running of the clocks associated with the moving receptor. Similarly, assuming that the emission frequency as observed by observers co-moving with the source to be correct, observers co-moving with the receptor expect to observe a particular frequency, whereas they observe a reduced frequency by the factor $\gamma$. The observers co-moving with the receptor interpret this result due to the slow running of clocks associated with the inertial frame co-moving with the source. Thus they reckon that the actual frequency of emission is lower by a factor $\gamma$ than that observed by the observers co-moving with the source. These results are consistent with the special relativity theory that two clocks in relative motion appear to run slow to each other. We have shown that this ratio remains invariant for all angles of emission and receipt of the light ray meaning that the mutual slow running of the clocks in relative motion depends only on the magnitude of the relative velocity, consistent with the theory of special relativity.

**Alternative Interpretation:** As the relativistic (observed) Doppler Effect is the geometric mean of the two classical Doppler Effects, one assuming that the speed of light is *c* with respect to the source and the other assuming that the speed of light is *c* with respect to the receptor, a common sense interpretation will be that neither of these assumptions are true but light travelled at an average speed that is in between the above two speeds. Maintaining the equivalence and isotropy of the inertial frames associated with the source and receptor and indeed all inertial frames, one may suggest that light travels at speed *c* with respect to material objects in the vicinity of those material objects (like the source of light and receptor). This resolution will maintain the equivalence of all inertial frames that are in relative motion with respect to each other. The other possible resolution can be a preferred inertial frame as envisaged originally by Lorentz wherein the moving objects actually contract along the direction of motion by a factor $(1-v^2/c^2)^{1/2}$ and moving clocks actually run slow by a factor $(1-v^2/c^2)^{1/2}$. This resolution will not hold all inertial frames to be equivalent.